\documentclass[11pt,english,epsf]{article}
\usepackage{amsmath}
\usepackage{amssymb}
\usepackage{amsfonts}
\usepackage{graphicx}
\usepackage{color}

\newtheorem{theorem}{Theorem}
\newtheorem{lemma}{Lemma}
\newcommand {\dfn} {\stackrel{\Delta} {=}}
\newcommand {\exe} {\stackrel{\cdot} {=}}
\newcommand {\lexe} {\stackrel{\cdot} {\le}}

\newcommand {\bs} {\mbox{\boldmath $s$}}

\newcommand {\bw} {\mbox{\boldmath $w$}}
\newcommand {\bx} {\mbox{\boldmath $x$}}
\newcommand {\by} {\mbox{\boldmath $y$}}

\newcommand {\bE} {\mbox{\boldmath $E$}}

\newcommand {\hP} {\hat{P}}
\newcommand {\hH} {\hat{H}}
\newcommand {\hI} {\hat{I}}

\newcommand {\bS} {\mbox{\boldmath $S$}}

\newcommand {\bW} {\mbox{\boldmath $W$}}
\newcommand {\bX} {\mbox{\boldmath $X$}}
\newcommand {\bY} {\mbox{\boldmath $Y$}}

\newcommand{\calA}{{\cal A}}

\newcommand{\calE}{{\cal E}}

\newcommand{\calG}{{\cal G}}

\newcommand{\calI}{{\cal I}}

\newcommand{\calS}{{\cal S}}
\newcommand{\calT}{{\cal T}}

\newcommand{\calW}{{\cal W}}
\newcommand{\calX}{{\cal X}}
\newcommand{\calY}{{\cal Y}}

\allowdisplaybreaks

\begin{document}
\thispagestyle{empty}
\title{False--Accept/False--Reject Trade--offs\\ in
Biometric Authentication Systems% Based on Secret Key Generation
%\thanks{This research was supported by my wife and kids.}
}
\author{Neri Merhav
%\thanks{
%Currently on sabbatical leave at HP Laboratories,
%1501 Page Mill Road, MS 3U-4, Palo Alto CA 94304, USA.}
}
\date{}
\maketitle

\begin{center}
The Andrew \& Erna Viterbi Faculty of Electrical Engineering\\
Technion - Israel Institute of Technology \\
Technion City, Haifa 32000, ISRAEL \\
E--mail: {\tt merhav@ee.technion.ac.il}\\
\end{center}
\vspace{1.5\baselineskip}
\setlength{\baselineskip}{1.5\baselineskip}

\begin{abstract}
Biometric authentication systems,
based on secret key generation, work as follows. In the enrollment
stage, an individual provides a biometric signal that is mapped into a secret
key and a helper message, the former being prepared to become
available to the system at a later time (for authentication),
and the latter is stored in a public database. When an authorized user
requests authentication, claiming his/her identity to be one of those of 
the subscribers,
he/she has to provide a biometric signal again, and then the system, which
retrieves also the helper message of the claimed subscriber, produces an
estimate of the secret
key, that is finally compared to the secret key of the claimed user. In case
of a match, the authentication request is approved, otherwise, it is rejected.

Evidently, there is an inherent tension between 
two desired, but conflicting, properties of
the helper message encoder: on the one hand, the encoding should be informative
enough concerning the identity of the real subscriber, in order to approve
him/her in the authentication stage,
but on the other hand, it
should not be too informative, as otherwise, unauthorized imposters could
easily fool the system and gain access. A good encoder should then trade
off the two kinds of errors: the false reject (FR) error and the false
accept (FA) error.

In this work, we investigate trade--offs between the random coding FR error exponent
and the best achievable FA error exponent. We compare two types of ensembles
of codes: fixed--rate codes and variable--rate codes, and we show that the
latter class of codes offers considerable improvement compared to the former.
In doing this, we characterize optimal rate functions for both types of
codes. We also examine the effect of privacy leakage constraints 
for both fixed--rate codes and variable--rate codes.
\end{abstract}


\newpage
\section{Introduction}

We consider a biometric authentication system
which is based on the one described in
\cite[Sections 2.2--2.6]{IW10}, and on the notion of secret key
generation and sharing of Maurer \cite{Maurer93}
and Ahlswede and Csisz\'ar \cite{AC93}, \cite{AC98}. In particular,
this system works in the following manner. In the enrollment
phase, a person that subscribes to the system, feeds it with
a biometric signal, $\bX =(X_1,X_2,\ldots,X_n)$. The system then responds by
generating (using its encoder) two outputs. The first
is a secret key, $\bS$, at rate $R_{\mbox{\tiny s}}$
and the second is a helper message, $\bW$,
at rate $R_{\mbox{\tiny w}}$. The secret key will be
used by the system later, at the authentication stage and the helper message
is saved in a database. When an authorized user (a subscriber)
wishes to sign in, claiming to be one of the subscribers that have already
enrolled, he/she is requested to provide again his/her biometric signal,
$\bY=(Y_1,\ldots,Y_n)$ (correlated to $\bX$, if indeed from the same
person, or independent, otherwise). The system then
retrieves the helper message $\bW$ of the claimed subscriber from the database,
and responds (using its decoder) by estimating the secret
key, $\hat{\bS}$ (based on $(\bY,\bW)$), and comparing it to that of
the claimed user, $\bS$. If $\hat{\bS}$ matches $\bS$, the
access to the system is approved, otherwise, it is denied.

In \cite[Sect.\ 2.3]{IW10}, the achievable region of pairs of rates $(R_{\mbox{\tiny s}},
R_{\mbox{\tiny w}})$ was established for the existence of authentication systems
where the following four quantities need to be arbitrarily small for large $n$: 
(i) the false--reject (FR)
probability, (ii) the false--accept (FA) probability,
(iii) the
privacy leakage,
$I(\bX;\bW)/n$, and (iv) the secrecy leakage, $I(\bS;\bW)/n$. Specifically, Theorem 2.1 of
\cite{IW10} asserts that when $(\bX,\bY)$ are drawn from a joint discrete memoryless
source (DMS), emitting independent copies of a pair of dependent random
variables, $(X,Y)\sim
P_{XY}$, the largest achievable key rate, $R_{\mbox{\tiny s}}$,
under the above constraints,
is given by $I(X;Y)$.
It then follows that $R_{\mbox{\tiny w}}$ must
lie between $H(X|Y)$
and $H(X)-R_{\mbox{\tiny s}}$,
where the lower limit is needed for
good identification of the legitimate subscriber
(small FR probability) as well as for achieving the minimum possible privacy leakage,
whereas the upper limit is
due to the secrecy leakage requirement. These limitations already
assure that $R_{\mbox{\tiny w}} < H(X)$, which in turn is necessary for keeping
the FA probability vanishingly small for large $n$. The achievability parts of
the corresponding coding theorems were proved in \cite{IW10} using random
binning, similarly as in classical Slepian--Wolf coding.

More recently, in \cite{p199} these results have been refined by characterizing
achievable exponential error bounds for the above performance metrics. In
particular, for a given rate pair $(R_{\mbox{\tiny s}},R_{\mbox{\tiny w}})$,
random coding error exponents and expurgated error exponents were found for the FR
probability, as well as sphere--packing bound, which is tight at a certain
region of the plane of $(R_{\mbox{\tiny s}},R_{\mbox{\tiny w}})$. For the FA
probability, the exact best achievable error exponent was characterized, and
finally, more refined upper bounds for privacy leakage and the secrecy
leakage were derived.

This paper is a further development of \cite{p199}, where the focus is on the
trade--off between the FA error exponent and the FR error exponent. In the
design of the helper message encoder, the following conflict arises: on the
one hand, it is desirable that the helper message $\bW$ would be informative
enough about $\bS$, such that in the presence of $\bY$, the identity of the
legitimate subscriber will be approved with high probability. But on the other
hand, it is also desired that in the absence of $\bY$, the helper message
would tell as little as possible about $\bS$, in order to make it difficult
for imposters to access the system. 

Indeed, the converse theorem in
\cite[Theorem 5]{p199} is based on the assumption that every type class of
source sequences $\{\bX\}$, is mapped, by the helper--message encoder, to as many different 
helper messages $\{\bW\}$ as possible, thus making it as close as possible to
be a one--to--one mapping,
or in other words, making $\bW$ is ``as
informative as possible'' about the source vector $\bX$, and hence also
about the secret key $\bS$, generated from $\bX$. This is good for achieving a small FR
probability (or, equivalently, a large FR error exponent), at the expense of
a limitation on the achievable FA exponent. In particular,
by relaxing the above described assumption,
and allowing smaller numbers of various helper messages for each source type
class, one may achieve better FA exponents, at the expense of worse FR
exponents. 

This raises the interesting question of achievable trade-offs
between the FA exponent and the FR exponent, which is similar, in spirit, to
the trade--off between the false alarm probability and the mis--detection
probability in the Neyman--Pearson scenario, where this trade--off is
traditionally encapsulated by the notion of receiver operating characteristics (ROC).
The difference, however, is that while in the Neyman--Pearson setting, this
trade--off is controlled by the choice of a detector (or more precisely, by the
choice of the threshold of the likelihood ratio detector), here we control the
trade-off via the choice of an encoder, in this case, the helper--message encoder.

To this end, we first derive an expression of the FR random coding error exponent
as a function of the desired FA error exponent for fixed--rate binning. This
is a relatively straightforward manipulation of the results of \cite{p199},
but it will serve as a reference result. The more interesting part, however, is about
extending the scope to the ensemble of variable--rate random binning codes,
whose {\it rate function} depends on the source vector only via its type
(similarly as in \cite{WM15} and
\cite{CHJL17}). The are two questions that arise in this context.
The first is: {\bf what are the optimal rate functions} of the secret--message and the
helper--message for maximizing the achievable FR exponent for a given FA error exponent? 
Upon finding such rate functions, the second question is: what is the
achievable FR error exponent as a function of the FA error exponent, and to
what extent does it improve relative to fixed--rate binning? We find an exact
formula for this function and demonstrate that the improvement may be rather
significant compared to fixed--rate binning. 
Finally, we examine the influence
of adding a constraint on the privacy leakage, in addition to the above
mentioned constraint on the FA error exponent for both fixed--rate codes and
variable--rate codes. 

On a technical note, it should be pointed out that while in \cite{p199}, the
error exponent expressions are provided in the Csisz\'ar--style formulation
(i.e., minimizations of certain functionals of information measures over
probability distributions), here we pass to Gallager--style formulations
(i.e., maximizations of functions of relatively few parameters). The reasons for
our interest in Gallager--style expressions are that they lend themselves more
conveniently to numerical calculations (see the discussion after Theorem
\ref{thm1} below, for more details), and that they may be
of independent interest on their own right.

The outline of the remaining part of this paper is as follows.
Section II establishes the notation conventions.
Section III provides a formal definition of the problem setting, 
then it gives some background (preliminaries), and finally, it describes the objectives.
Section IV provides a preparatory step of deriving the optimal rate functions
that maximize an 
expression of the achievable FR error exponent for a given FA error exponent, in both
fixed--rate and variable--rate regimes. In Section V, we derive the FA error
exponents of both fixed-- and variable--rate codes as functions the
prescribed FA error exponent. Finally, in Section VI, we examine the effect
of a constraint on the privacy leakage.

\section*{II. Notation Conventions}

Throughout the paper, random variables will be denoted by capital
letters, specific values they may take will be denoted by the
corresponding lower case letters, and their alphabets
will be denoted by calligraphic letters. Random
vectors and their realizations will be denoted,
respectively, by capital letters and the corresponding lower case letters,
both in the bold face font. Their alphabets will be superscripted by their
dimensions. For example, the random vector $\bX=(X_1,\ldots,X_n)$, ($n$ --
positive integer) may take a specific vector value $\bx=(x_1,\ldots,x_n)$
in $\calX^n$, the $n$--th order Cartesian power of $\calX$, which is
the alphabet of each component of this vector.
Sources and channels will be denoted by the letter $P$ or $Q$,
subscripted by the names of the relevant random variables/vectors and their
conditionings, if applicable, following the standard notation conventions,
e.g., $Q_X$, $P_{Y|X}$, and so on. When there is no room for ambiguity, these
subscripts will be omitted.
The probability of an event $\calG$ will be denoted by $\mbox{Pr}\{\calG\}$,
and the expectation
operator with respect to (w.r.t.) a probability distribution $P$ will be
denoted by
$\bE_P\{\cdot\}$. Again, the subscript will be omitted if the underlying
probability distribution is clear from the context.
The entropy of a generic distribution $Q$ on $\calX$ will be denoted by
$H_Q(X)$. For two
positive sequences $a_n$ and $b_n$, the notation $a_n\exe b_n$ will
stand for equality in the exponential scale, that is,
$\lim_{n\to\infty}\frac{1}{n}\log \frac{a_n}{b_n}=0$. Similarly,
$a_n\lexe b_n$ means that
$\limsup_{n\to\infty}\frac{1}{n}\log \frac{a_n}{b_n}\le 0$, and so on.
The indicator function
of an event $\calG$ will be denoted by $\calI\{\calG\}$. The notation $[x]_+$
will stand for $\max\{0,x\}$.

The empirical distribution of a sequence $\bx\in\calX^n$, which will be
denoted by $\hat{P}_{\bx}$, is the vector of relative frequencies
$\hat{P}_{\bx}(x)$
of each symbol $x\in\calX$ in $\bx$.
The type class of $\bx\in\calX^n$, denoted $\calT(\hP_{\bx})$, is the set of
all vectors $\bx'$
with $\hat{P}_{\bx'}=\hat{P}_{\bx}$.
Information measures associated with empirical distributions
will be denoted with `hats' and will be subscripted by the sequences from
which they are induced. For example, the entropy associated with
$\hat{P}_{\bx}$, which is the empirical entropy of $\bx$, will be denoted by
$\hat{H}_{\bx}(X)$.
Similar conventions will apply to the joint empirical
distribution, the joint type class, the conditional empirical distributions
and the conditional type classes associated with pairs (and multiples) of
sequences of length $n$.
Accordingly, $\hP_{\bx\by}$ will be the joint empirical
distribution of $(\bx,\by)=\{(x_i,y_i)\}_{i=1}^n$,
and $\calT(\hP_{\bx\by})$ will denote
the joint type class of $(\bx,\by)$. Similarly, $\calT(\hP_{\bx|\by}|\by)$
will stand for
the conditional type class of $\bx$ given
$\by$, $\hH_{\bx\by}(X,Y)$
will designate the empirical joint entropy of $\bx$
and $\by$,
$\hH_{\bx\by}(X|Y)$ will be the empirical conditional entropy,
$\hI_{\bx\by}(X;Y)$ will
denote empirical mutual information, and so on.
We will also use similar rules of notation in the context of
a generic distribution, $Q_{XY}$ (or $Q$, for short): we use
$\calT(Q_X)$ for the type class of sequences with empirical distribution
$Q_X$,
$H_Q(X)$ -- for the corresponding empirical entropy,
$\calT(Q_{XY})$ -- for the joint type class x,
$T(Q_{X|Y}|\by)$ -- for the conditional type class of $\bx$ given $\by$,
$H_Q(X,Y)$ -- for the joint empirical entropy,
$H_Q(X|Y)$ -- for the conditional empirical entropy,
$I_Q(X;Y)$ -- for the empirical mutual information, and so on.
We will also use the customary notation for the weighted divergence,
\begin{equation}
D(Q_{Y|X}\|P_{Y|X}|Q_X)=\sum_{x\in\calX}Q_X(x)\sum_{y\in\calY}Q_{Y|X}(y|x)\log
\frac{Q_{Y|X}(y|x)}{P_{Y|X}(y|x)}.
\end{equation}

\section*{III. Problem Setting, Preliminaries and Objectives}

\subsection*{A. Problem Setting}

The problem setting is similar to the one in \cite{p199}, but with a few small
differences, mainly related to the fact that here, in contrast to \cite{p199},
we allow variable--rate binning codes.

Consider the following system model
for biometric identification. An {\it enrollment source
sequence}, $\bx=(x_1,\ldots,x_n)$, that is a realization of the random vector
$\bX=(X_1,\ldots,X_n)$,
that emerges from a discrete memoryless source (DMS),
$P_X$, with a finite alphabet $\calX$, is fed into an {\it enrollment
encoder}, $\calE$, that generates two
outputs: a secret key, $\bs$ (a realization of a random variable
$\bS$), and a helper
message, $\bw$ (a realization of $\bW$), both taking values in finite
alphabets, $\calS_n$ and $\calW_n$, respectively. In the {\it fixed--rate
regime}, $\calS_n=\{1,2,\ldots,e^{nR_{\mbox{\tiny s}}}\}$ and
$\calW_n=\{1,2,\ldots,e^{nR_{\mbox{\tiny w}}}\}$ (assuming that
$e^{nR_{\mbox{\tiny s}}}$ and $e^{nR_{\mbox{\tiny w}}}$ are integers), where
$R_{\mbox{\tiny s}}$ is the {\it secret--key rate}, and
$R_{\mbox{\tiny w}}$ is the {\it helper--message rate}. In the {\it variable--rate
regime}, we allow both rates to depend on the type $Q_X$ of the given input
vector $\bx$. In particular, in the variable rate regime, each 
$\bx\in\calT(Q_X)$ is mapped, by the secret--key
encoder and by the helper--message encoder, into
$\bs\in\calS_n(Q_X)=\{0,1,\ldots,e^{nR_{\mbox{\tiny s}}(Q_X)}\}$ and
$\bw\in\calW_n(Q_X)=\{0,1,\ldots,e^{nR_{\mbox{\tiny w}}(Q_X)}\}$, respectively, where
$R_{\mbox{\tiny s}}(Q_X)$ and $R_{\mbox{\tiny w}}(Q_X)$, henceforth referred
to as {\it rate functions}, are given continuous
functions of $Q_X$. These encodings designate the enrollment stage.

Since the fixed--rate case is obviously a special case of
the variable--rate case, our description will henceforth relate
to the variable--rate case, with the understanding that in the fixed--rate case,
$R_{\mbox{\tiny s}}(Q_X)$ and $R_{\mbox{\tiny w}}(Q_X)$ are just constants,
denoted $R_{\mbox{\tiny s}}$ and $R_{\mbox{\tiny w}}$, independent of $Q_X$.

As in \cite{IW10}, we consider the ensemble of enrollment encoders,
$\{\calE\}$, generated by {\it random binning}, where
for each source vector $\bx\in\calX$, one selects independently at random,
both a secret key and a helper message, under the uniform distributions across
$\calS_n(Q_X)$ and $\calW_n(Q_X)$, respectively. We denote by
$\bw=f(\bx)$ and
$\bs=g(\bx)$, the randomly selected bin assignments for both outputs.

The {\it authentication decoder}, $\calA$, which is aware of the randomly
selected encoder, $\calE$, is fed by two inputs: the helper message $\bw$ and
an {\it authentication source sequence}, $\by=(y_1,\ldots,y_n)$ (a realization of
$\bY=(Y_1,\ldots,Y_n)$), that is produced at
the output of a discrete memoryless channel (DMC),
$P_{Y|X}$, with a finite output alphabet $\calY$, that is fed by $\bx$.
The output of the authentication decoder
is $\hat{\bs}=U(\by,\bw)$ (a realization of $\hat{\bS}$), which is an estimate
(possibly, randomized) of the
secret key, $\bs$. If $\hat{\bs}=\bs$, access to the system is granted,
otherwise, it is denied. This decoding operation stands for the authentication
stage.

The optimal estimator of $\bs$, based on $(\by,\bw)$, in the sense of
minimum FR probability, $\mbox{Pr}\{\hat{\bS}\ne\bS\}$, is
the maximum a posteriori probability (MAP) estimator, given by
\begin{equation}
\hat{\bs}_{\mbox{\tiny MAP}}=U(\by,\bw)\dfn\mbox{arg}\max_{\bs}
P(\bs,\bw|\by)=
\mbox{arg}\max_{\bs}
\sum_{\bx\in\calX^n}P(\bx|\by)\cdot\calI\{f(\bx)=\bw\}\cdot\calI\{g(\bx)=\bs\},
\end{equation}
where $P(\bx|\by)$ (shorthand notation for $P_{\bX|\bY}(\bx|\by)$)
is the posterior probability of $\bX=\bx$ given $\bY=\by$,
that is induced by the product distribution, $P_{XY}$ (and the
subscript $XY$ will sometimes be suppressed for simplicity,
when there is no risk of compromising clarity).

As in \cite{p199}, here too, we consider the framework of 
generalized stochastic likelihood decoders (GLDs)
\cite{p187}, \cite{SMF15}, \cite{SCP14}, \cite{YAG13}, where the decoder
randomly selects its output $\hat{\bs}$ according to the posterior
distribution
\begin{equation}
\label{gld}
\tilde{P}(\bs|\by,\bw)=\frac{\sum_{\bx\in\calX^n}
\exp\{na(\hP_{\bx\by})\}\cdot\calI\{f(\bx)=\bw\}\cdot\calI\{g(\bx)=\bs\}}
{\sum_{\bx\in\calX^n}
\exp\{na(\hP_{\bx\by})\}\cdot\calI\{f(\bx)=\bw\}},
\end{equation}
where the function $a(\cdot)$, which will be referred to as the {\it decoding
metric}, is any
continuous function of the joint empirical
distribution $\hP_{\bx\by}$. 
As explained in \cite{p199}, as well as in earlier
studies, the motivation for considering GLDs
is that they provide a unified framework for examining a
large variety of decoders.
For example, with
\begin{equation}
\label{ordinary1}
a(\hP_{\bx\by})=\sum_{x\in\calX}\sum_{y\in\calY}\hP_{\bx\by}(x,y)\ln
P_{X|Y}(x|y),
\end{equation}
we have the ordinary likelihood decoder \cite{SMF15},
\cite{SCP14}, \cite{YAG13}. For
\begin{equation}
\label{ordinary2}
a(\hP_{\bx\by})=\beta\sum_{x\in\calX}\sum_{y\in\calY}\hP_{\bx\by}(x,y)\ln
P_{X|Y}(x|y),
\end{equation}
$\beta > 0$ being a parameter, we
extend this to a parametric family of decoders.
In particular,
$\beta\to\infty$ leads to the ordinary MAP decoder, $\hat{\bs}_{\mbox{\tiny
MAP}}$. Other choices of $a(\cdot)$ are associated with mismatched metrics,
\begin{equation}
\label{mismatched}
a(\hP_{\bx\by})=\sum_{x\in\calX}\sum_{y\in\calY}\hP_{\bx\by}(x,y)\ln
P^{\prime}(x|y),
\end{equation}
$P^{\prime}$ being different from $P_{X|Y}$, and
\begin{equation}
\label{minimumentropy}
a(\hP_{\bx\by})=-\beta\hH_{\bx\by}(X|Y),
\end{equation}
which for $\beta\to\infty$,
tends to the universal minimum entropy decoder. When
$a(\hP_{\bx\by})=\beta\cdot\alpha(\hP_{\bx\by})$, $\alpha(\cdot)$ being an arbitrary
function and $\beta\to\infty$, we end up with
Csisz\'ar's $\alpha$--decoder \cite{CK81}.

An illegal user (imposter),
who claims for a given legal identity,
does not
have the correlated biometric data $\by$, and so, the best he/she can do is to
estimate $\bs$
based on $\bw$, and then
forge any fake biometric data $\tilde{\by}$,
which together with $\bw$, would cause the
decoder to output this estimate of $\bs$. More precisely, the imposter first
estimates $\bs$ according to
\begin{equation}
\label{imposter}
\tilde{\bs}=V(\bw)\dfn\mbox{arg}\max_{\bs} P(\bs|\bw)=
\mbox{arg}\max_{\bs}
\sum_{\bx\in\calX^n}P(\bx)\cdot\calI\{f(\bx)=\bw\}\cdot\calI\{g(\bx)=\bs\},
\end{equation}
and then generates any $\tilde{\by}\in\calY^n$
such that $U(\tilde{\by},\bw)=\tilde{\bs}$, and uses it as the
biometric signal for authentication.

\subsection*{B. Preliminaries}

In \cite[Theorems 1, 4, and 5]{p199}, the following two results (among others) were
derived for fixed--rate binning at rates $R_{\mbox{\tiny w}}$ and $R_{\mbox{\tiny
s}}$: the best
achievable FA exponent is given by
\begin{equation}
\label{p199a}
E_{\mbox{\tiny
FA}}(R_{\mbox{\tiny w}},R_{\mbox{\tiny s}})=
\min_{Q_X}[D(Q_X\|P_X)+\min\{R_{\mbox{\tiny s}},[H_Q(X)-R_{\mbox{\tiny w}}]_+\}],
\end{equation}
and the random coding FR exponent is given by,
\begin{equation}
\label{p199b}
E_{\mbox{\tiny FR}}
(R_{\mbox{\tiny w}})=\min_{Q_{X_0Y}}\{D(Q_{X_0Y}\|P_{XY})+E(R_{\mbox{\tiny
w}},Q_{X_0Y})\},
\end{equation}
where
\begin{equation}
\label{p199c}
E(R_{\mbox{\tiny w}},Q_{X_0Y})
=\min_{Q_{X|Y}}[R_{\mbox{\tiny w}}-H_Q(X|Y)+[a(Q_{X_0Y})-a(Q_{XY})]_+]_+.
\end{equation}
As shown in \cite[eq.\ (12)]{p199}, for the decoding metric $a(Q)=-H_Q(X|Y)$, 
eq.\ (\ref{p199c}) simplifies to
\begin{equation}
\label{p199d}
E(R_{\mbox{\tiny w}},Q_{X_0Y})=[R_{\mbox{\tiny w}}-H_Q(X_0|Y)]_+,
\end{equation}
which is equivalent to the error exponent expression corresponding to optimal MAP
decoding, $\hat{\bs}_{\mbox{\tiny MAP}}$ (i.e., eq.\ (\ref{ordinary2}) with
$\beta\to\infty$).

\subsection*{C. Objectives}

As described in the Introduction, our first objective is to derive the FR error
exponent as a function of the prescribed FA error exponent, henceforth
referred to as the {\it FR--FA trade-off function},
for fixed--rate
codes with optimal rate functions and optimal decoding metrics. 
This FR--FA trade-off function will be derived from eqs.\
(\ref{p199a})--(\ref{p199c}). The more interesting goal would then be to extend the
scope to variable--rate codes, derive optimal rate functions,
$R_{\mbox{\tiny w}}^*(Q_X)$ and
$R_{\mbox{\tiny s}}^*(Q_X)$, then use them to obtain the FR--FA trade--off
function for variable--rate codes together with their own
optimal decoding metrics, and finally, compare to the trade-off function of
fixed--rate codes. 

Another objective is to examine the effect of imposing a privacy leakage
constraint in addition to the FA error exponent constraint. This will be
carried out in both the
fixed--rate regime and the variable--rate regime. 

\section*{IV. Optimal Rate Functions and Decoding Metrics}

We begin by deriving optimal rate functions for both fixed--rate codes and
variable--rate codes.

\subsection*{A. Fixed--Rate Codes}

For fixed--rate codes, the following lemma establishes the optimal
helper--message rate, $R_{\mbox{\tiny w}}$, and secret key rate,
$R_{\mbox{\tiny s}}$, for a given value, $E_0 > 0$, of the FA error exponent, 
$E_{\mbox{\tiny FA}}$.

\begin{lemma}
\label{lemma1}
Necessary and sufficient conditions for the existence of fixed--rate codes
that achieve $E_{\mbox{\tiny FA}}(R_{\mbox{\tiny
w}},R_{\mbox{\tiny s}})\ge E_0$ are:
\begin{eqnarray}
R_{\mbox{\tiny s}}&\ge&E_0,\label{rsdemand}\\
R_{\mbox{\tiny w}}&\le&R_{\mbox{\tiny w}}^*(E_0)\nonumber\\
&\dfn&\min_{\{Q_X:~D(Q_X\|P_X)\le E_0\}}
\bE_Q\log\frac{1}{P_X(X)}- E_0\nonumber\\
&=&\sup_{\lambda\ge 0}\left\{-\lambda\ln\left(\sum_{x\in\calX}
[P_X(x)]^{1+1/\lambda}\right)-(1+\lambda)E_0\right\}.\label{rwdemand}
\end{eqnarray}
\end{lemma}

Note that the requirement $R_{\mbox{\tiny s}}\ge E_0$ is quite intuitive,
because even a blind guess of $\bS$ may succeed with probability of
$e^{-nR_{\mbox{\tiny s}}}$. It was shown in \cite{WI12} that the best
achievable FA exponent is given in turn by $I(X;Y)$. This is coherent with the
result \cite[Theorem 2.1]{IW10} that $I(X;Y)$ is also an achievable upper bound on
$R_{\mbox{\tiny s}}$.

\noindent
{\it Proof of Lemma \ref{lemma1}.}
From eq.\ (\ref{p199a}), it is immediately seen that the statement, $E_{\mbox{\tiny
FA}}(R_{\mbox{\tiny
w}},R_{\mbox{\tiny s}})\ge E_0$, is equivalent to the statement 
\begin{equation}
\forall~Q_X~~~D(Q_X\|P_X)+\min\{R_{\mbox{\tiny s}},[H_Q(X)-R_{\mbox{\tiny
w}}]_+\}\ge E_0,
\end{equation}
which in turn is equivalent to the two simultaneous statements,
\begin{eqnarray}
& &\forall~Q_X~~~R_{\mbox{\tiny s}}\ge E_0-D(Q_X\|P_X)\\
& &\forall~Q_X~~~[H_Q(X)-R_{\mbox{\tiny w}}]_+\ge E_0-D(Q_X\|P_X)
\end{eqnarray}
The former happens if and only if $R_{\mbox{\tiny s}}\ge E_0$, which is 
eq.\ (\ref{rsdemand}). As for the latter, for $D(Q_X\|P_X)\ge E_0$, the
r.h.s.\ is non--positive, whereas the l.h.s.\ is non--negative, and so, there is no
limitation on $R_{\mbox{\tiny w}}$, which is associated with the region
$\{Q_X:~D(Q_X\|P_X)\ge E_0\}$. For $D(Q_X\|P_X) < E_0$, on the other hand,
we must have $H_Q(X)-R_{\mbox{\tiny w}}\ge E_0-D(Q_X\|P_X)$,
or equivalently,
\begin{equation}
R_{\mbox{\tiny w}}\le H_Q(X)+D(Q_X\|P_X)-E_0\equiv
\bE_Q\ln\frac{1}{P_X(X)}-E_0,
\end{equation}
for every $Q_X$ such that $D(Q_X\|P_X) < E_0$.
This, in turn, is equivalent to the requirement given in the first two lines
of eq.\ (\ref{rwdemand}). The third line of (\ref{rwdemand}) is obtained as
follows:
\begin{eqnarray}
R_{\mbox{\tiny w}}^*(E_0)
&=&\min_{\{Q_X:~D(Q_X\|P_X)\le E_0\}}
\bE_Q\log\frac{1}{P_X(X)}- E_0\nonumber\\
&=&\min_{Q_X}\sup_{\lambda\ge
0}\left\{\bE_Q\log\frac{1}{P_X(X)}+\lambda[D(Q_X\|P_X)-E_0]-E_0\right\}\nonumber\\
&=&\sup_{\lambda\ge
0}\min_{Q_X}\left\{\bE_Q\log\frac{1}{P_X(X)}+\lambda[D(Q_X\|P_X)-E_0]-E_0\right\}\nonumber\\
&=&\sup_{\lambda\ge 0}\left\{-\lambda\ln\left(\sum_{x\in\calX}
[P_X(x)]^{1+1/\lambda}\right)-(1+\lambda)E_0\right\},
\end{eqnarray}
where the third equality is follows from convexity in $Q_X$ and concavity
(in fact, affinity) in $\lambda$.

\subsection*{B. Variable--Rate Codes}

For variable--rate codes, we have the following lemma, which sets the stage
for optimal rate functions.

\begin{lemma}
\label{lemma2}
Necessary and sufficient conditions for the existence of variable--rate codes
that achieve FA error exponent at least as large as $E_0$ are:
\begin{eqnarray}
R_{\mbox{\tiny s}}(Q_X)&\ge& R_{\mbox{\tiny s}}^*(Q_X,E_0)\dfn E_0-D(Q_X\|P_X),\\
R_{\mbox{\tiny w}}(Q_X)&\le&R_{\mbox{\tiny w}}^*(Q_X,E_0)\nonumber\\
&\dfn&\left\{\begin{array}{ll}
\bE_Q\log\frac{1}{P_X(X)}- E_0 & D(Q_X\|P_X)< E_0\\
\infty & D(Q_X\|P_X)\ge E_0.\end{array}\right.
\end{eqnarray}
\end{lemma}

Observe that $R_{\mbox{\tiny s}}^*(Q_X,E_0)+R_{\mbox{\tiny
w}}^*(Q_X,E_0)=H_Q(X)$ for all $Q_X$ with $D(Q_X\|P_X)< E_0$, which 
roughly speaking, means that
the mapping from $\bx$ to $(\bs,\bw)$ is one--to--one within each type,
$\calT(Q_X)$.\\

\noindent
{\it Proof of Lemma \ref{lemma2}.}
Eq.\ (\ref{p199a}) easily extends to the variable--rate case, by simply
substituting $R_{\mbox{\tiny s}}(Q_X)$ and $R_{\mbox{\tiny w}}(Q_X)$ instead
of $R_{\mbox{\tiny s}}$ and $R_{\mbox{\tiny w}}$, respectively. Therefore, the
same reasoning as in the proof of Lemma \ref{lemma1} applies in the
variable--rate setting considered here as well, except that
now, there is no need for optimization (maximization, in the case of
$R_{\mbox{\tiny s}}$, and minimization, in the case of $R_{\mbox{\tiny w}}$),
as the binning rates are allowed to depend on the type, $Q_X$.

\section*{V. FR--FA Trade-off Functions}

In this section, we characterize FR--FA trade-off functions for both
the random coding ensembles of both fixed--rate and variable--rate codes.

\subsection*{A. Fixed--Rate Codes}

According to eq.\
(\ref{p199b}), the random coding FR exponent depends on $R_{\mbox{\tiny
w}})$ only, and it is a
monotonically, non--decreasing function of this variable. Thus, 
the best one can do with fixed--rate
codes is to use the highest allowable binning rate, which is $R_{\mbox{\tiny w}}^*$.
Therefore, if we denote the fixed--rate FR--FA trade-off function by $E_{\mbox{\tiny
FR}}^{\mbox{\tiny f}}[E_0]$ (where the superscript ``f'' stands for
``fixed--rate''), 
we have the following expression for the optimal decoding metric,
$a(Q)=-H_Q(X|Y)$ (see eq.\ (\ref{p199d})),
\begin{equation}
\label{frexponent-csiszar}
E_{\mbox{\tiny FR}}^{\mbox{\tiny f}}[E_0]=E_{\mbox{\tiny FR}}(R_{\mbox{\tiny w}}^*(E_0))=
\min_{Q_{XY}}\{D(Q_{XY}\|P_{XY})+[R_{\mbox{\tiny w}}^*(E_0)-H_Q(X|Y)]_+\},
\end{equation}
which is also well known to be the Csisz\'ar--style formula for 
the error exponent associated with ordinary MAP decoding 
(of the full source vector $\bX$, rather than just 
the secret key $\bS$) for a random Slepian--Wolf
code (see, e.g., \cite[eqs.\ (7), (19)]{CHJL17}). 

As mentioned already in the Introduction, in this paper, we are also
be interested in Gallager--style forms, since they are more convenient to work
with when it comes to numerical calculations.
The Gallager--style form of eq.\ (\ref{frexponent-csiszar}) is well known
\cite{Gallager76}, \cite[p.\ 9]{CHJL17} to be
\begin{equation}
\label{frexponent-fixed-gallager}
E_{\mbox{\tiny FR}}^{\mbox{\tiny f}}[E_0]=\max_{0\le\rho\le
1}\left\{-\ln\left[\sum_{y\in\calY}\left(\sum_{x\in\calX}
[P_{XY}(x,y)]^{1/(1+\rho)}\right)^{1+\rho}\right]
+\rho R_{\mbox{\tiny w}}^*(E_0)\right\}.
\end{equation}
Upon substituting the expression of $R_{\mbox{\tiny w}}^*(E_0)$ (see eq.\
(\ref{rwdemand})), we finally obtain
\begin{eqnarray}
E_{\mbox{\tiny FR}}^{\mbox{\tiny f}}[E_0]&=&
\max_{0\le \rho \le 1}\sup_{\lambda \ge 0}\left\{
-\ln\left[\sum_{y\in\calY}\left(
\sum_{x\in\calX}[P_{XY}(x,y)]^{1/(1+\rho)}\right)^{1+\rho}
\right]-\right.\nonumber\\
& &\left.\rho\lambda \ln\left(\sum_{x\in\calX}[P_X(x)]^{1+1/\lambda}\right)-
\rho(1+\lambda)E_0\right\}.
\end{eqnarray}

\subsection*{B. Variable--Rate Codes}

We now derive a FR--FA trade-off function for the variable--rate case, which
will be denoted by $E_{\mbox{\tiny FR}}^{\mbox{\tiny v}}[E_0]$. 

The analysis associated with variable--rate codes is somewhat more complicated than with
fixed--rate codes. Note that since the alphabets, $\calW_n(Q_X)$ and
$\calS_n(Q_X)$, of $\bw$ and $\bs$, respectively, depend
on the type class, $Q_X$, of the source sequence, a given $\bw$
can be generated only
by types for which $e^{nR_{\mbox{\tiny w}}(Q_x)}$ is at least as large 
as the numerical index\footnote{By ``numerical index'', we mean the
integer corresponding to the location of $\bw$ within $\calW_n(Q_X)$,
which is a number between $1$ and $e^{nR_{\mbox{\tiny w}}(Q_x)}$.} of $\bw$,
and a similar comment applies to $\bs$. This means that in the generalized
posterior, defined in (\ref{gld}), the summations 
over the source vectors, $\{\bx\}$, at both the
numerator and the denominator, should now be limited only to members of the
type classes that support the given $\bs$ and $\bw$. Consequently, some
modifications in the analysis of \cite[Proof of Theorem 1]{p199} should be carried out
in the variable--rate case.\footnote{In particlar, the maximizations over
$\{Q_{X|Y}\}$, in eqs.\ (19), (22) of \cite{p199}, and the minimizations in
eq.\ (23) therein, should be limited only to types $\{Q_{X|Y}\}$ that pertain
to $\{Q_X\}$ that support the given $\bw$.} It is easy to see, however, that
a valid lower bound\footnote{While the exact FR random coding 
error exponent can be derived in
principle, it
is much more complicated to use than this lower bound. Nonetheless, as we show
in the sequel, if even this lower bound would yield a significant improvement
relative to fixed--rate codes (as we demonstrate in the sequel), 
then a--fortiori, this would also be the case with the exact FR error exponent.
Also, this lower bound is tight for all $\{\bw\}$ whose numerical index is
a sub--exponential function of $n$ (and then supported by essentially all types
$\{Q_X\}$).} 
to the resulting FR error exponent is obtained if one
simply substitutes $R_{\mbox{\tiny w}}(Q_X)$ instead of the fixed $R_{\mbox{\tiny w}}$
of eqs.\ (\ref{p199b})--(\ref{p199d}). In other words, we will use the
expression,
\begin{eqnarray}
\label{frbound}
E_{\mbox{\tiny FR}}(R_{\mbox{\tiny w}}(\cdot),Q_{X_0Y})&=&
\min_{Q_{X_0Y}}\{D(Q_{X_0Y}\|P_{X_0Y})+\nonumber\\
& &\min_{Q_{X|Y}}[R_{\mbox{\tiny w}}(Q_X)-H_Q(X|Y)+[a(Q_{X_0})-a(Q_{XY})]_+]_+\}.
\end{eqnarray}
Once again,
since the FR exponent of eqs.\ (\ref{p199b})--(\ref{p199d}) 
is a monotonically non--decreasing function of the helper--message
rate, the best we can do in terms of 
this expression, is to let $R_{\mbox{\tiny w}}(Q_X)$ saturate its maximum
allowed value, $R_{\mbox{\tiny
w}}^*(Q_X,E_0)$, as given in Lemma 2.

Having adopted eq.\ (\ref{frbound}) as our figure of merit,
the following point is important as well:
the universal decoding metric 
$a(Q_{XY})=-H_Q(X|Y)$, that we have used above for fixed--rate codes,
is no longer equivalent to that of MAP decoding (and hence no longer optimal) for
variable--rate codes. For a given rate function, $R_{\mbox{\tiny w}}(Q_X)$, 
the following decoding metric should be used instead in
order to obtain the same random coding FR exponent as in MAP decoding:
\begin{equation}
a(Q_{XY})= R_{\mbox{\tiny w}}(Q_X)-H_Q(X|Y).
\end{equation}
In this case, referring to eqs.\ (10) and (12) of \cite{p199}, we have
\begin{eqnarray}
E(R_{\mbox{\tiny w}}(\cdot),Q_{X_0Y})&=&
\inf_{Q_{X|Y}}[R_{\mbox{\tiny w}}(Q_X)-H_Q(X|Y)+[a(Q_{X_0Y})-a(Q_{XY})]_+]_+\nonumber\\
&=&\inf_{Q_{X|Y}}[R_{\mbox{\tiny w}}(Q_X)-H_Q(X|Y)+[R_{\mbox{\tiny w}}
(Q_{X_0})-H_Q(X_0|Y)-\nonumber\\
& &\{R_{\mbox{\tiny w}}(Q_X)-H_Q(X|Y)\}]_+]_+\nonumber\\
&=&\inf_{Q_{X|Y}}[\max\{R_{\mbox{\tiny w}}(Q_X)-H_Q(X|Y),R_{\mbox{\tiny w}}
(Q_{X_0})-H_Q(X_0|Y)\}]_+]_+\nonumber\\
&=&[R_{\mbox{\tiny w}}(Q_{X_0})-H_Q(X_0|Y)]_+\nonumber\\
&\ge&\min\{[R_{\mbox{\tiny w}}(Q_{X})-H_Q(X|Y)]_+:~\bE_Q\ln P(X|Y)\ge \bE_Q\ln
P(X_0|Y)\},
\end{eqnarray}
where the last line corresponds to (pairwise) errors pertaining to the MAP decoder
(for ordinary Slepian--Wolf decoding), and hence the optimality.
Now, for $R_{\mbox{\tiny w}}(Q_X)=R_{\mbox{\tiny w}}^*(Q_X,E_0)$,
we can present the above bound to the 
FR error exponent (omitting the subscript $0$ of $X_0$,
which is no longer needed):
\begin{equation}
\label{frexponent-variable-csiszar}
E_{\mbox{\tiny FR}}^{\mbox{\tiny v}}[E_0]=
\min_{\{Q_{XY}:~D(Q_{X}\|P_X)\le E_0
\}}\left\{D(Q_{XY}\|P_{XY})+\left[\bE_Q\ln\frac{1}{P_X(X)}-E_0-H_Q(X|Y)\right]_+\right\}.
\end{equation}
This is the Csisz\'ar--style formula of the corresponding FR--FA trade-off function for
variable--rate codes. The following theorem provides the Gallager--style form
of the same function.
\begin{theorem}
\label{thm1}
The variable--rate FR--FA trade-off function
(\ref{frexponent-variable-csiszar}) can also be presented as
\begin{eqnarray}
\label{frexponent-variable-gallager}
E_{\mbox{\tiny FR}}^{\mbox{\tiny v}}[E_0]&=&
\max_{0\le\lambda\le 1}\sup_{\rho\ge
0}\max_W\left\{-\ln\left(\sum_{y\in\calW}\left[\sum_{x\in\calX}[P_{XY}(x,y)P_X(x)^{\rho+\lambda}
W(x)^{-\rho}]^{1/(1+\lambda)}\right]^{1+\lambda}\right)-\right.\nonumber\\
& &\left.(\rho+\lambda)E_0\right\},
\end{eqnarray}
where the maximum over $W$ is taken over the simplex of probability
distributions over $\calX$, i.e., $W(x)\ge 0$ for all $x\in\calX$ and
$\sum_{x\in\calX}W(x)=1$.
\end{theorem}

Before turning to the proof of Theorem \ref{thm1}, we pause to demonstrate it
and to discuss some
aspects, consequences and extensions of this theorem.\\

\noindent
{\bf Optimization issues.}
First, note that the formula (\ref{frexponent-variable-gallager}) 
involves optimization over a probability
distribution $W$ in addition to the parameters $\rho$ and $\lambda$, namely, a
total of $|\calX|+1$ parameters. This number is never larger (and in most cases,
considerably smaller) than the
$|\calX|\cdot|\calY|-1$ parameters that are associated with the minimization
over $Q_{XY}$ in the Csisz\'ar--style formula of eq.\
(\ref{frexponent-variable-csiszar}). Moreover, since the Gallager--style
formula (\ref{frexponent-variable-gallager}) involves only maximization, any
arbitrary choice of $\lambda$, $\rho$ and $W$ in their allowed ranges, would
yield a valid lower bound (a guarantee) on the achievable FR error exponent. This is
different from the situation with the Csisz\'ar--style formula, which involves
minimization, and hence allows no such privilege: one {\bf must} carry out the
minimization in order to obtain the achievable FR error exponent. One drawback
of the Gallager--style formula is that the range of maximization over the
parameter $\rho$ is infinite. In practical numerical calculations, 
however, one can initially limit the range
to an interval of the form $[0,\rho_0]$, 
and then gradually enlarge $\rho_0$ up to the point
where no further increase in $\rho_0$ improves on the resulting maximum.

A few words are in order concerning the
maximization over $W$, which is a relatively computationally demanding step,
especially for a large source alphabet. First, note that
in situations with a sufficient degree of symmetry, 
the optimal $W$ turns out to be the uniform
distribution, a fact that saves the optimization numerically. This
turns out to be the case if $P_X$
is uniform and the $|\calY|$ probability vectors
$\{P_{X|Y=y}(x|y),~x\in\calX\}$, are all
permutations, $\{\pi_y\}$, of one such vector, that form a group (w.r.t.\
compositions of permutations), such that $(1/|\calY|)\sum_y
\pi_y^{-1}[W]=U$, where $U$ designates the uniform distribution over $\calX$.
This happens, for instance, when $P_X=U$ and $P_{X|Y}$ is a
modulo--additive channel. To see why this is true, we define
\begin{eqnarray}
f(W)&=&\inf_{Q_{XY}}\sum_{y\in\calY}
Q_Y(y)\ln\frac{Q_Y(y)}{P_Y(y)}+\sum_{y\in\calY}Q_Y(y)\sum_xQ_{X|Y}(x|y)\left[\ln
\frac{Q_{X|Y}(x|y)}{P_{X|Y}(x|y)}+\right.\nonumber\\
& &\left.(\rho+\lambda)\ln\frac{1}{P_X(x)}+
\lambda\ln Q_{X|Y}(x|y)
+\rho\ln W(x)\right],
\end{eqnarray}
which we show\footnote{See the proof of Theorem \ref{thm1} in the sequel.} 
to be equal to
\begin{equation}
f(W)=-\ln\left(\sum_{y\in\calY}\left[\sum_{x\in\calX}[P_{XY}(x,y)P_X(x)^{\rho+\lambda}
W(x)^{-\rho}]^{1/(1+\lambda)}\right]^{1+\lambda}\right).
\end{equation}
From the first representation of $f$, it is easy to see that it is concave in
$W$. From the second representation and the assumed symmetry, it is easy to
see that for every $W$ and every $y\in\calY$, $f(\pi_y[W])=f(W)$. Thus,
\begin{equation}
f(W)=\frac{1}{|\calY|}\sum_{y\in\calY}f(\pi_y[W])\le
f\left(\frac{1}{|\calY|}\sum_{y\in\calY}\pi_y[W]\right)=f(U),
\end{equation}
which means that the optimal $W$ is uniform. 
In the general case, the optimization over $W$ is a convex program,
and so, there are standard solvers that can
handle this problem more efficiently than a brute--force exhaustive search.\\

\begin{figure}[h!t!b!]
\centering
\includegraphics[width=8.5cm, height=8.5cm]{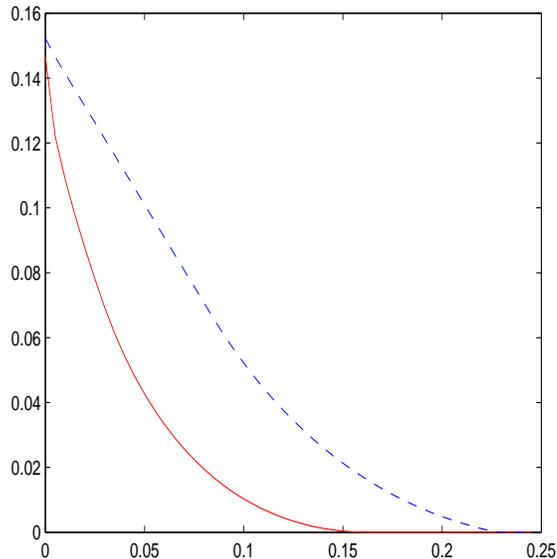}
\caption{Graphs of $E_{\mbox{\tiny FR}}^{\mbox{\tiny f}}[E_0]
$ (solid red curve) and $E_{\mbox{\tiny FR}}^{\mbox{\tiny
v}}[E_0]$ (dashed blue curve) for the double binary source,
defined by
$P_{XY}(0,0)=0.32$,
$P_{XY}(0,1)=0.08$,
$P_{XY}(1,0)=0.06$, and
$P_{XY}(1,1)=0.54$.}
%These graphs are constructed as follows: For fixed-rate coding and
%variable-rare coding, use the
%MATLAB functions maxf(efa) and maxv(efa), respectively, where efa is the
%desired FA exponent.
%The functions are written for a double binary source (namely, both X and Y
%are binary).
\label{graphs}
\end{figure}

\noindent
{\bf Example.}
In order to demonstrate the advantage of variable--rate codes relative to
fixed--rate codes in terms of the FR--FA trade-off, we now provide a 
numerical example. Consider the case of a double binary source with alphabets
$\calX=\calY=\{0,1\}$, and
joint probabilities given by
$P_{XY}(0,0)=0.32$,
$P_{XY}(0,1)=0.08$,
$P_{XY}(1,0)=0.06$, and
$P_{XY}(1,1)=0.54$.
Fig.\ \ref{graphs} displays the two FR--FA trade-off functions,
$E_{\mbox{\tiny FR}}^{\mbox{\tiny f}}[E_0]$ and
$E_{\mbox{\tiny FR}}^{\mbox{\tiny v}}[E_0]$, for this source.
As can be seen, the gap between these two functions is
rather considerable, which means that variable--rate codes with optimal rate
functions, are significantly better in terms of these trade--offs.
This example is quite representative in the sense that
other examples (with different source probabilities)
yielded qualitatively similar results.

\noindent
{\bf Variable--rate codes do not always improve on fixed--rate codes.}
It should be kept in mind, however, that there are situations where
variable--rate codes offer no improvement over fixed--rate codes, i.e., they
might have exactly the same FR--FA trade-off function in some cases. 
One such example is the case
where the source $X$ has a uniform distribution. In this case, the optimal
rate function for variable--rate codes turns out to be $R_{\mbox{\tiny
w}}^*(Q_X,E_0)=\ln|\calX|-E_0$ (whenever $D(Q_X\|P_X) < E_0$),
which is independent of $Q_X$, and hence is a fixed--rate anyway. So 
unless the dominant type $Q_X$ happens
to fall in the region $\{Q_X:~D(Q_X\|P_X)\ge E_0\}$, when the source is
uniform, variable--rate codes
cannot offer any improvement beyond the performance of fixed--rate codes.

Another aspect is associated with the decoding metric. Even
for a general source, $P_{XY}$, 
if one uses a random variable--rate code, but decodes
it using the decoding metric function of fixed--rate codes,
$a(Q_{XY})=-H_Q(X|Y)$,
instead of the optimal decoding metric for variable--rate codes, $a(Q_{XY})=
R_{\mbox{\tiny w}}^*(Q_X,E_0)-H_Q(X|Y)$, then the resulting FR--FA trade--off
function turns out to be exactly the same as with fixed--rate codes.

\noindent
{\bf Mismatched decoding.}
Our results can be extended to apply to a mismatched decoding metric,
$a(Q_{XY})=\bE_Q\ln P'(X|Y)$ (see eq.\ (\ref{mismatched})), for an arbitrary
$P'$, using the same
techniques. The resulting fixed--rate Gallager--style FR--FA trade-off function
would then be
\begin{eqnarray}
E_{\mbox{\tiny FR}}^{\mbox{\tiny f}}[E_0]&=&
\max_{0\le s\le 1}
\max_{0\le t\le
s}\left\{-\ln\left(\sum_{y\in\calY}\left[\left(\sum_{x\in\calX}[P'(x|y)]^{t/s}\right)^s\cdot\left(
\sum_{x'\in\calX}\frac{P_{XY}(x',y)}{[P'(x'|y)]^t}\right)\right]\right)+\right.\nonumber\\
& &\left.sR_{\mbox{\tiny
w}}^*(E_0)\right\},
\end{eqnarray}
where $R_{\mbox{\tiny w}}^*(E_0)$ is as defined in Lemma \ref{lemma1}. The
corresponding variable--rate rae-off function (which cannot be smaller), is given by
\begin{eqnarray}
E_{\mbox{\tiny FR}}^{\mbox{\tiny v}}[E_0]&=&
\sup_{\lambda\ge 0}\max_{0\le s\le 1}
\max_{0\le t\le
s}\max_W\left\{-\ln\left(\sum_{y\in\calY}\left[\left(\sum_{x\in\calX}[P'(x|y)]^{t/s}
[P_X(x)]^{1+\lambda/s}[W(x)]^{-\lambda/s}\right)^s\times\right.\right.\right.\nonumber\\
& &\left.\left.\left.\left(
\sum_{x'\in\calX}\frac{P_{XY}(x',y)}{[P'(x'|y)]^t}\right)\right]\right)-%\right.\nonumber\\
%& &\left.
(\lambda+s)E_0\right\}.
\end{eqnarray}

\noindent
{\bf Expurgated exponents.}
In \cite{p199}, expurgated FR exponents were also derived, and so, in
principle, one could
carry out similar analyses for trade--offs between expurgated exponents and
the best achievable FA error exponents. 
We have not pursued such derivations in this work since they
are significantly more complicated, but it is anticipated 
that similar conclusions would apply concerning
the advantage of variable--rate codes over
fixed--rate codes. In this context, it should be emphasized that 
one of our important messages in this work is not only that
variable--rate codes are better than fixed--rate codes, but that moreover, we
characterize the {\it optimal rate functions}, independently of the type
of FR error exponents being considered (random coding exponents, expurgated
exponents, sphere--packing exponent \cite[Theorem 3]{p199} at a certain rate
region, etc.)
since their derivation stems from the FA error exponent,
which has an exact characterization \cite[Section V]{p199}.

The remaining part of this section is devoted to the proof of Theorem
\ref{thm1}.\\

\noindent
{\it Proof of Theorem \ref{thm1}.}
Throughout this proof we will make frequent use of the minimax theorem, based
on convexity--concavity arguments. We will also use repeatedly the fact that
$$\min_Q[D(Q\|P)+\bE_Qf(X)]=-\ln\left[\sum_x P(x)e^{-f(x)}\right].$$
Now,
\begin{eqnarray}
E_{\mbox{\tiny FR}}^{\mbox{\tiny v}}[E_0]
&=&\inf_{Q_{XY}}
\sup_{0\le\lambda\le 1}\sup_{\rho\ge
0}\left\{D(Q_{XY}\|P_{XY})+\lambda\left[\bE_Q\ln\frac{1}{P_X(X)}-E_0
-H_Q(X|Y)\right]+.\right.\nonumber\\
& &\left.\rho\left[D(Q_X\|P_X)-E_0\right]\right\}\nonumber\\
&=&\sup_{0\le\lambda\le 1}\sup_{\rho\ge
0}\inf_{Q_{XY}}\left\{D(Q_{XY}\|P_{XY})+\lambda\left[\bE_Q\ln\frac{1}{P_X(X)}-E_0
-H_Q(X|Y)\right]+\right.\nonumber\\
& &\left.\rho\left[D(Q_X\|P_X)-E_0\right]\right\}\nonumber\\
&=&\sup_{0\le\lambda\le 1}\sup_{\rho\ge
0}\inf_{Q_Y}\left\{D(Q_Y\|P_Y)+\inf_{Q_{X|Y}}\left(D(Q_{X|Y}\|P_{X|Y}|Q_Y)+\right.\right.\nonumber\\
& &\left.\left.\lambda\left[\bE_Q\ln\frac{1}{P_X(X)}-E_0
-H_Q(X|Y)\right]
+\rho\left[D(Q_X\|P_X)-E_0\right]\right)\right\}\nonumber\\
&=&\sup_{0\le\lambda\le 1}\sup_{\rho\ge
0}\inf_{Q_Y}\left\{D(Q_Y\|P_Y)+\inf_{Q_{X|Y}}\sum_{y\in\calY}Q_Y(y)\sum_xQ_{X|Y}(x|y)\left[\ln
\frac{Q_{X|Y}(x|y)}{P_{X|Y}(x|y)}+\right.\right.\nonumber\\
& &\left.\left.(\rho+\lambda)\ln\frac{1}{P_X(x)}+
\lambda\ln Q_{X|Y}(x|y)
+\rho\ln Q_X(x)\right]-(\rho+\lambda)E_0\right\}.
\end{eqnarray}
Consider first the inner--most minimization over $\{Q_{X|Y}\}$: 
\begin{eqnarray}
& &\inf_{Q_{X|Y}}\sum_{y\in\calY}Q_Y(y)\sum_{x\in\calX}Q_{X|Y}(x|y)\left[\ln
\frac{Q_{X|Y}(x|y)}{P_{X|Y}(x|y)}+
(\rho+\lambda)\ln\frac{1}{P_X(x)}+
\lambda\ln Q_{X|Y}(x|y)
+\rho\ln Q_X(x)\right]\nonumber\\
&=&\inf_{Q_{X|Y}}\sup_W\sum_{y\in\calY}Q_Y(y)\sum_{x\in\calX}Q_{X|Y}(x|y)\left[\ln
\frac{Q_{X|Y}(x|y)}{P_{X|Y}(x|y)}+
(\rho+\lambda)\ln\frac{1}{P_X(x)}+
\lambda\ln Q_{X|Y}(x|y)
+\rho\ln W(x)\right]\nonumber\\
&=&\sup_W\sum_{y\in\calY}Q_Y(y)\inf_{Q_{X|Y}}\sum_{x\in\calX}Q_{X|Y}(x|y)\left[\ln
\frac{Q_{X|Y}(x|y)}{P_{X|Y}(x|y)}+
(\rho+\lambda)\ln\frac{1}{P_X(x)}+
\lambda\ln Q_{X|Y}(x|y)
+\rho\ln W(x)\right]\nonumber\\
&=&(1+\lambda)\sup_W\sum_{y\in\calY}Q_Y(y)\inf_{Q_{X|Y}}\sum_{x\in\calX}Q_{X|Y}(x|y)
\ln\frac{Q_{X|Y}(x|y)}
{[P_{X|Y}(x|y)P_X(x)^{\rho+\lambda}W(x)^{-\rho}]^{1/(1+\lambda)}}\nonumber\\
&=&-(1+\lambda)\inf_W\sum_yQ(y)\ln\left(\sum_x[P(x|y)P(x)^{\rho+\lambda}W(x)^{-\rho}]^{1/(1+\lambda)}\right),
\end{eqnarray}
which, after the minimization over $Q_Y$, becomes
\begin{eqnarray}
& &\inf_{Q_Y}\left[D(Q_Y\|P_Y)-(1+\lambda)\inf_W\sum_{y\in\calY}Q_Y(y)
\ln\left(\sum_{x\in\calX}[P_{X|Y}(x|y)P_X(x)^{\rho+\lambda}
W(x)^{-\rho}]^{1/(1+\lambda)}\right)\right]\nonumber\\
&=&-\inf_W\ln\left(\sum_{y\in\calY}\left[\sum_{x\in\calX}[P_{XY}(x,y)P_X(x)^{\rho+\lambda}
W(x)^{-\rho}]^{1/(1+\lambda)}\right]^{1+\lambda}\right).
\end{eqnarray}
It follows that
\begin{eqnarray}
E_{\mbox{\tiny FR}}^{\mbox{\tiny v}}[E_0]&=&
\max_{0\le\lambda\le 1}\sup_{\rho\ge
0}\sup_W\left\{-\ln\left(\sum_{y\in\calY}\left[\sum_{x\in\calX}[P_{XY}(x,y)P_X(x)^{\rho+\lambda}
W(x)^{-\rho}]^{1/(1+\lambda)}\right]^{1+\lambda}\right)-(\rho+\lambda)E_0
\right\}.\nonumber
\end{eqnarray}
This completes the proof of Theorem \ref{thm1}.$\Box$

\section*{VI. Privacy Leakage}

The privacy leakage of the system is defined as $I(\bX;\bW)$.
Since $\bW$ is 
a deterministic function of $\bX$, we have $I(\bX;\bW)=H(\bW)=-\bE\ln P(\bW)$.

Consider first variable--rate codes.
For a given, arbitrarily small $\epsilon > 0$, let
\begin{equation}
\calT_\epsilon(P_X)=\bigcap_{\{Q_X:~D(Q_X\|P_X)\le\epsilon\}}\calT(Q_X),
\end{equation}
Let $\bw$ be a helper message whose numerical index
does not exceed $|\calT_\epsilon(P_X)|$.
Now, for the typical code,
\begin{eqnarray}
P(\bw)&=&\sum_{\bx}P(\bx)\calI\{f(\bx)=\bw)\}\nonumber\\
&\ge&\sum_{\bx\in\calT_\epsilon(P_X)}P(\bx)\calI\{f(\bx)=\bw)\}\nonumber\\
&\ge&\sum_{\bx\in\calT_\epsilon(P_X)}2^{-n[H_P(X)+\epsilon]}\cdot
N(\calT_\epsilon(P_X),\bw),
\end{eqnarray}
where
$N(\calT_\epsilon(P_X),\bw)=|\calT_\epsilon(P_X)\cap\{\bx:~f(\bx)=\bw\}|$,
which is the sum of independent
binomial random variables (one for every $Q_X$ in the $\epsilon$--neighborhood
of $P_X$)
with a total of $|\calT_\epsilon(P_X)|$ trials and success
rate of at least $\exp_2[-n\max\{R_{\mbox{\tiny
w}}(Q_X):~D(Q_X\|P_X)\le\epsilon\}]$. Assuming that $R_{\mbox{\tiny
w}}(P_X) < H_P(X)$, then if $R_{\mbox{\tiny w}}(\cdot)$ is continuous and
$\epsilon$ is very small, we have
that for the typical code,
$N(\calT_\epsilon(P_X),\bw)$ is of the exponential order of
$2^{n[H_P(X)-R_{\mbox{\tiny w}}(P_X)-\delta(\epsilon)]}$, where
$\delta(\epsilon)$ is some function that tends to zero as $\epsilon$ tends to
zero. Thus, for the typical code,
$P(\bw)$ (for $\bw$ whose numerical index is 
in the designated range) is essentially lower bounded by
the exponential order of $2^{-nR_{\mbox{\tiny w}}(P_X)}$.
The probability that $\bw$ falls in the complementary set, where its numerical
index exceeds $|\calT_\epsilon(P_X)|$ is upper bounded by the probability that
$\bX\notin \calT_\epsilon(P_X)$, which tends to zero as $n\to\infty$.
Consequently, this set
of (a--typical) $\{\bw\}$ has an asymptotically vanishing contribution to the
entropy of $\bW$. It follows that $H(\bW)$ is essentially upper bounded by
$nR_{\mbox{\tiny w}}(P_X)$, and for the optimal rate function, this means
$n[H_P(X)-E_0]$. This means that if $H(\bW)$ is required to be smaller than $nH_0$, for
some given $H_0$, this can be the case only if $H_0 \ge H_P(X)-E_0$.
Since the best achievable $E_0$ is $I_P(X;Y)$ (see \cite[Theorem 1]{WI12}), then
the smallest achievable $H_0$ is about $H_P(X|Y)$, which is coherent with
\cite[Proposition 2.4]{IW10}.

We therefore summarize our conclusion as follows: it is possible to 
comply with both a privacy leakage
constraint, $H_0$, and a FA error exponent constraint, $E_0$, as long as
$H_0\ge H_P(X)-E_0$. As intuition suggests, there is no real tension
between the requirements on privacy leakage and FA error exponent, as they both
work in the same direction of making it hard for imposters. As the former
becomes more restrictive, the latter can be made more relaxed, and vice versa.

In the fixed--rate case the situation is somewhat different. 
Let $N(Q_X,\bw)$ be the number of $\bx$--vectors of type $Q_X$ for which
$f(\bx)=\bw$, which is a binomial random variable with
$|\calT(Q_X)|\exe e^{nH_Q(X)}$ trials and success rate of $e^{-nR_{\mbox{\tiny
w}}}$. Now, for the typical code,
\begin{eqnarray}
P(\bw)&=&\sum_{\bx}P(\bx)\calI\{f(\bx)=\bw)\}\nonumber\\
&=&\sum_{Q_X}P(\bx)\bigg|_{\bx\in\calT(Q_X)}\cdot N(Q_X,\bw)\nonumber\\
&\exe&\sum_{\{Q_X:~H(Q_X)\ge R_{\mbox{\tiny
w}}\}}e^{-n[H(Q_X)+D(Q_X\|P_X)]}\cdot
e^{n[H(Q_X)-R_{\mbox{\tiny w}}]}\nonumber\\
&\exe&\exp\left\{-n\left[R_{\mbox{\tiny w}}+\min_{\{Q_X:~H(Q_X)\ge
R_w\}}D(Q_X\|P_X)\right]\right\},
\end{eqnarray}
independently of $\bw$, and so, for the typical code
\begin{equation}
H(\bW) \approx n\left[R_{\mbox{\tiny w}}+\min_{\{Q_X:~H(Q_X)\ge R_{\mbox{\tiny
w}}\}}D(Q_X\|P_X)\right].
\end{equation}
If we impose the constraint $H(\bW)\le nH_0$ (in addition to the FA
constraint),
for some prescribed constant $H_0$, then
this constraint is equivalent to the requirement that there would exist
at least one pmf $Q_X$ for which $R_{\mbox{\tiny w}}\le
\min\{H(Q_X),H_0-D(Q_X\|P_X)\}$, or equivalently,
\begin{eqnarray}
R_{\mbox{\tiny w}}&\le&\max_{Q_X}\min\{H_Q(X),H_0-D(Q_X\|P_X)\}\nonumber\\
&=&\max_{Q_X}\min_{0\le s\le 1}\{(1-s)H_Q(X)+s[H_0-D(Q_X\|P_X)]\}\nonumber\\
&=&\min_{0\le s\le 1}\max_{Q_X}\{(1-s)H_Q(X)+s[H_0-D(Q_X\|P_X)]\}\nonumber\\
&=&\min_{0\le s\le 1}\max_{Q}\sum_{x\in\calX}Q_X(x)\left\{-(1-s)\ln
Q_X(x)+s\ln\frac{P_X(x)}{Q_X(x)}+sH_0\right\}\nonumber\\
&=&\min_{0\le s\le 1}\max_{Q_X}\sum_{x\in\calX}Q_X(x)\left\{\ln\frac{P_X^s(x)}
{Q_X(x)}+sH_0\right\}\nonumber\\
&=&\min_{0\le s\le 1}\left\{\ln\left[\sum_{x\in\calX} P_X^s(x)\right]+sH_0
\right\}\nonumber\\
&\dfn&R_{\mbox{\tiny w}}^{*}(H_0).
\end{eqnarray}
Of course, if both a privacy constaint and an FA error exponent constraint are
imposed at the same time, then for fixed--rate codes,
$R_{\mbox{\tiny w}}$ is not allowed to exceed
$\min\{R_{\mbox{\tiny w}}^*(E_0),
R_{\mbox{\tiny w}}^{*}(H_0)\}$. 

%\section*{Acknowledgements}

%\section*{Appendix}
%\renewcommand{\theequation}{A.\arabic{equation}}
%    \setcounter{equation}{0}

\end{document}